\documentclass[aps,prd,preprintnumbers,showpacs]{revtex4}
\usepackage[dvips]{graphicx}
\begin{document}

%
%

\eprint{Nisho-2-2019}
\title{Chiral Symmetry Breaking by Monopole Condensation}
\author{Aiichi Iwazaki}
\affiliation{Nishogakusha University,\\ 
6-16 Sanbancho Chiyoda-ku Tokyo 102-8336, Japan.}   
\date{Jan. 17, 2019}
\begin{abstract}
Under the assumption of Abelian dominance in QCD,
we show that either color charge or chirality of quark is not conserved, when the low energy massless quark collides with QCD monopole.
Because the color charge is conserved in reality, the chirality is not conserved.  
We show by using chiral anomaly 
that chiral non symmetric quark pair production takes place when a classical color charge is putted in a vacuum
with monopole condensation, 
while chiral symmetric pair production takes place in a vacuum with no monopole condensation.  
Our results indicate that the chiral symmetry is broken by the monopole condensation.
\end{abstract}
\hspace*{0.3cm}
\hspace*{1cm}


\maketitle

It has been shown with lattice gauge theories\cite{karsch} that quark confinement and chiral symmetry breaking simultaneously arises in SU(3) gauge theory
with massless quark color triplets.
That is, the transition temperature between confinement and deconfinement phases 
almost coincides with the transition temperature 
between chiral symmetric and antisymmetric phases. Although extensive studies\cite{suga,miyamura,w,fuku,gat,sch,iritani} have been performed,
the explicit connection between the confinement and the chiral symmetry breaking has still not been
clear.  
The confinement
is caused by the monopole condensation\cite{nambu,man,thooft} in the analysis with the use of maximal Abelian gauge\cite{mag}. 
On the other hand, the chiral symmetry
breaking is caused by the chiral condensation of quark-antiquark pair. It has been discussed that the condensate arises with 
instanton effects according to chiral anomaly. No intimate relation between
the monopole condensate and the chiral condensate was found, although there were 
numerical evidences\cite{miyamura} of the relation. But,  it has recently been discussed\cite{hasegawa,iwazaki3} that
the monopoles induce the chiral condensate.

In this paper we show using chiral anomaly that when an external color charge is put in a vacuum with the monopole condensation,
the vacuum expectation value $\langle dQ_5/dt \rangle $ 
does not vanish,
while it vanishes when the vacuum has no monopole condensation. Here $dQ_5/dt$ denotes the time derivative of the chirality $Q_5$;
$Q_5=N_R-N_L$ where $N_R$ ( $N_L$ ) denotes the number of the right ( left ) handed quarks.
That is, $\langle dQ_5/dt\rangle $ 
represents the pair production of the massless quarks under the background electric field of the external charge. 
Therefore, the chiral non symmetric pair production of the massless quarks takes place in the vacuum with monopole condensation.
It indicates that the chiral symmetry is broken by the monopole condensation. 
In contrast, the chiral non symmetric pair production does not take place
when there are no magnetic excitations in vacuum. 
The vacuum with the monopole condensation is characterized such 
that it is not eigenstate of the magnetic charge oprerator $\hat{Q}_m$; $\hat{Q}_m|v\rangle \neq q_m|v\rangle$.
Hereafter,
we take SU(2) gauge theory for simplicity.

The positive charged fermion
produced moves to the direction of the electric field, while the negative charged fermion does to the opposite direction of the electric field.
Their spins can take arbitrary directions parallel or anti parallel to the electric field. There are no favorable directions. Therefore, the production
is chiral symmetric; $\langle dQ_5/dt\rangle=0$. 
On the other hand, the favorable directions of the spins spontaneously arise in the vacuum with the monopole condensation,
even if there are no magnetic fields. 
This leads to the spontaneous chiral symmetry breaking, $\langle dQ_5/dt\rangle\neq 0$.

\vspace{0.1cm}
First, we explain that either the charge or the chirality of a quark 
is not conserved in the low energy massless quark scattering with a Dirac monopole.
We need to impose an appropriate boundary condition on the quark at the location of the monopole, which
determines the conserved quantity.

Our concern is 
massless quark doublet $(\begin{array}{l}\Psi_+ \\ \Psi_-\end{array} )$ satisfying Dirac equation in the background gauge fields of the monopole,

\begin{equation}
\label{1}
\gamma_{\mu}(i\partial^{\mu}\mp \frac{g}{2}A^{\mu})\Psi_{\pm}=0
\end{equation}
where the gauge potentials $A^{\mu}$ denotes a Dirac monopole
given by 

\begin{equation}
\label{2}
A_{\phi}=g_m(1-\cos(\theta)), \quad A_0=A_r=A_{\theta}=0
\end{equation} 
where $\vec{A}\cdot d\vec{x}=A_rdr+A_{\theta}d\theta+A_{\phi}d\phi$ with polar coordinates $r,\theta$ and $\phi=\arctan(y/x)$.
$g_m$ denotes a magnetic charge with which magnetic field is given by $\vec{B}=g_m\vec{r}/r^3$.
The magnetic charge satisfies the Dirac quantization condition $g_mg=n/2$ with integer $n$ where $g$ denotes the U(1) gauge coupling.
Hereafter, we assume the monopoles with the magnetic charge $g_m=1/2g$.

\vspace{0.1cm}
The quark doublet coupled with the Dirac monopole arises
in SU(2) gauge theory under the assumption of the Abelian dominance\cite{iwa,suzuki}.
The maximal Abelian group is described by the diagonal component ( $\sigma_3$ ) of the SU(2) gauge fields. 
The Dirac monopole is represented by the Abelian gauge fields. 
Thus, the quark doublet $q=(q^+,q^-)$ carry the charges $\pm g/2$ associated with the diagonal component.

We will explain why either the charge or the chirality is not conserved when the quark collides with the monopole. 
As is well known, the conserved angular momentum\cite{coleman} of the quark under the magnetic monopole located at $r=0$ 
is given by $\vec{J}=\vec{L}+\vec{S}\mp gg_m\vec{r}/r$,
where $\vec{L}$ ( $\vec{S}$ ) denotes orbital ( spin ) angular momentum. The last term is peculiar to the particle under the background field
of the monopole. Owing to the term we can show that either the charge or the chirality is not conserved in the scattering.
In order to see it we note the conserved quantity $\vec{J}\cdot\vec{r}=\vec{S}\cdot\vec{r}\pm gg_m r$.
When the chirality ( or helicity $\sim \vec{p}\cdot\vec{S}/|\vec{p}||\vec{S}|$ ) is conserved, the spin must flip $\vec{S}\to -\vec{S}$ after the scattering 
because the momentum flips after the scattering; $\vec{p}\to -\vec{p}$.
Then, the charge must flip $g\to -g$ because of the conservation of $\vec{J}\cdot\vec{r}$, i.e.
 $\Delta(\vec{J}\cdot\vec{r})=\Delta(\vec{S}\cdot\vec{r})+\Delta(gg_mr)=0$. ( $\Delta(Q)$ denotes the change of the value $Q$ after the scattering. ) 
On the other hand, when the charge is conserved ( $0=\Delta(\vec{J}\cdot\vec{r})=\Delta(\vec{S}\cdot\vec{r}$) ),
the chirality $\vec{p}\cdot\vec{S}/|\vec{p}||\vec{S}|$ must flip because the spin does not flip $\vec{S}\to \vec{S}$.
In this way, either the charge or the chirality conservation is lost in the scattering.

To appropriately define the scatterings, it has been discussed\cite{kazama} 
that we need to imposed a boundary condition for the quarks at the location of the monopoles.
It is either of charged conserved but chirality non conserved boundary condition or chirality conserved but charge non conserved one.
The charge conservation is strictly guaranteed by the gauge symmetry.
Therefore, 
we need to impose the boundary conditions $q^{\pm}_R(r=0)=q^{\pm}_L(r=0)$ at the location of the monopole, 
which breaks the chiral symmetry.

We give another explanation of the chiral non conservation in the monopole quark scattering using chiral anomaly.
The anomaly equation describing the evolution of the chirality $Q_5$ is given by

\begin{equation}
\label{3}
\frac{dQ_5}{dt}=\frac{g^2}{4\pi^2}\int d^3r \vec{E}\cdot\vec{B}=\frac{g^2}{4\pi^2}\int d^3r \vec{E}\cdot\frac{g_m\vec{r}}{r^3}
=\frac{g^2}{4\pi^2}\int d^3r \frac{g(\vec{r}-\vec{x}(t))}{4\pi|\vec{r}-\vec{x}(t)|^3}\cdot\frac{g_m\vec{r}}{r^3}=\frac{g^3g_m}{4\pi^2|\vec{x}(t)|},
\end{equation}
where the electric field $\vec{E}=g(\vec{r}-\vec{x}(t))/(4\pi|\vec{r}-\vec{x}(t)|^3)$ is produced by a quark located at $\vec{x}(t)$.
The magnetic field $\vec{B}=g_m\vec{r}/r^3$ is produced by a monopole located at $\vec{r}=0$.
The anomaly equation describes how the chirality of the classical quark changes with time $t$, depending on its coordinate $\vec{x}(t)$. 
Here, we consider the scattering of the quark such that 
it goes from $\vec{x}(t=-\infty)=-\infty $ to $\vec{x}(t=\infty)=+\infty$,
passing $\vec{x}(t=0)=\vec{x}_0\neq 0$ at $t=0$. 
Then, when the quark does not flip its color charge after the scattering, the quantity $dQ_5/dt$ does not change its sign. Thus,
$Q_5(+\infty)-Q_5(-\infty)=\int_{-\infty}^{+\infty}dt\, dQ_5/dt$ does not vanish. 
The chirality is not conserved, when the charge is conserved.

We would like to mention the another meaning of the anomaly equation.  
The equation shows the quantum production of the chirality when a classical charged particle is put in a vacuum with the monopole.
Actually, it has been shown\cite{iwazaki} 
that the anomaly equation can describe quark pair production under the classical homogeneous fields $\vec{E}$ and $\vec{B}$.
The fields are produced by high energy heavy ion collisions.
The result is coincident with the one obtained by the analysis in the standard Schwinger mechanism. 
In our case we have a classical charged particle and a monopole which produce the electric field $\frac{g(\vec{r}-\vec{x}(t))}{4\pi|\vec{r}-\vec{x}(t)|^3}$
and the magnetic field $\frac{g_m\vec{r}}{r^3}$.
Then, the equation shows that the chiral non symmetric
pair production takes place under these external fields.

\vspace{0.2cm}
Now, we show using the anomaly equation 
that $\langle dQ_5/dt \rangle\neq 0$ when an external charge is put in the vacuum with the monopole condensation, 
while $\langle dQ_5/dt \rangle=0$ without the monopole condensation.
The nonvanishing of $\langle dQ_5/dt \rangle$ implies that the chiral non symmetric pair production of the massless quarks arises.

When there are monopoles with their magnetic charges $g_m\eta_i$ at $\vec{x}_i$ ( $ i=1,,,$ ) with $\eta_i=\pm 1$, the anomaly equation is given by

\begin{equation}
\frac{dQ_5(\vec{x})}{dt}=\sum_{i=1,,,}\frac{g^3g_m\eta_i }{4\pi^2|\vec{x}-\vec{x}_i|}=\int d^3y \frac{g^3\rho_m(\vec{y})}{4\pi^2|\vec{x}-\vec{y}|} \quad \mbox{with} \quad
\rho_m(\vec{y})\equiv \sum_{i=1,,,}\eta_ig_m\delta(\vec{y}-\vec{x}_i),
\end{equation}  
where $\rho_m$ denotes the magnetic charge density of the monopoles.

In order to discuss the quantum effects of the monopoles, 
we replace the classical magnetic charge density $\rho_m$ with the quantum operator $\hat{\rho}_m$ of the magnetic charge.
For instance, it is given such that    
$\hat{\rho}_m=g_m\Phi^{\dagger}iD_t\Phi+h.c.$ in terms of monopole field $\Phi$.
Because we wish to obtain the vacuum expectation value $\langle dQ_5/dt\rangle$,
we estimate it in the following

\begin{equation}
\label{12}
\lim_{|\vec{x}|\to \infty}\langle \frac{dQ_5(\vec{x})}{dt}\frac{dQ_5(0)}{dt} \rangle
=\lim_{|\vec{x}|\to \infty}\int d^3yd^3y' \frac{(g^3)^2\langle\hat{\rho}_m(\vec{y})\hat{\rho}_m(\vec{y'})\rangle}{(4\pi^2)^2|\vec{x}-\vec{y}||\vec{y}'|}
=\langle dQ_5/dt\rangle^2, 
\end{equation}
where the expectation value is taken in the vacuum of the monopoles.
Noting the translational invariance, we rewrite the equation(\ref{12}) with the use of the function 
$f(\frac{\vec{y}-\vec{y}'}{\sqrt{2}})\equiv\langle \hat{\rho}_m(\vec{y})\hat{\rho}_m(\vec{y}')\rangle$
and the variables $\vec{y}_{\pm}\equiv(\vec{y}\pm\vec{y}')/\sqrt{2}$,

\begin{eqnarray}
&&\lim_{|\vec{x}|\to \infty}\langle \frac{dQ_5(\vec{x})}{dt}\frac{dQ_5(0)}{dt} \rangle=
\lim_{|\vec{x}|\to \infty}\int d^3y_{+}d^3y_{-} \frac{2(g^3)^2f(\vec{y}_{-})}{(4\pi^2)^2|\vec{y}_{+}+2\vec{y}_{-}-\sqrt{2}\vec{x}||\vec{y}_{+}|} \nonumber \\
&=& \lim_{|\vec{x}|\to \infty}4\pi(g^3)^2\int_0^{\infty} dy_{+}\int d^3y_{-} 
\frac{\exp(-y_{+}/L)\Big(y_{+}+|2\vec{y}_{-}-\sqrt{2}\vec{x}|-|y_{+}-|2\vec{y}_{-}-\sqrt{2}\vec{x}||\Big)f(\vec{y}_{-})}{(4\pi^2)^2|2\vec{y}_{-}-\sqrt{2}\vec{x}|} \nonumber \\
&=& \lim_{|\vec{x}|\to \infty}L^2(g^3)^2 \int d^3y_{-} 
\frac{\Big(1-\exp(-|2\vec{y}_{-}-\sqrt{2}\vec{x}|/L)\Big)f(\vec{y}_{-})}{2\pi^3|2\vec{y}_{-}-\sqrt{2}\vec{x}|} \nonumber \\
&\simeq& \frac{L(g^3)^2}{2\pi^3}\int d^3y_{-} f(\vec{y}_{-})=\frac{L(g^3)^2}{2\pi^3}\int d^3y_{-}\langle \hat{\rho}_m(\sqrt{2}\vec{y}_{-})\hat{\rho}_m(0)\rangle 
=\frac{L(g^3)^2}{4\sqrt{2}\pi^3} \langle Q_m\,\hat{\rho}_m(0)\rangle ,
\end{eqnarray}
with the magnetic charge $\hat{Q}_m\equiv \int d^3x \hat{\rho}_m(\vec{x})$,
where we have introduced a cut off $L$ in the integration of $y_{+}$ and have taken a limit $L\to \infty$ before taking the limit $|\vec{x}|\to \infty$.
Therefore, the vacuum expectation value $\langle dQ_5/dt\rangle$ is given by

\begin{equation}
\label{14}
 \langle v|\frac{dQ_5}{dt}|v\rangle=\pm \sqrt{\frac{L(g^3)^2}{4\pi^3\sqrt{2}} \langle v|\hat{Q}_m\,\hat{\rho}_m(0)|v \rangle},
\end{equation}
where we denote the vacuum as $|v\rangle$.

The equation (\ref{14}) implies that $\langle v=0|dQ_5/dt|v=0\rangle=0$ when  $\langle v=0|\Phi|v=0\rangle= 0$,
while $\langle v|dQ_5/dt|v\rangle \neq 0$ when the monopole condenses $\langle v|\Phi|v\rangle=v \neq 0$.
Actually, when the monopole does not condense, the vacuum is an eigenstate of the magnetic charge $\hat{Q}_m$. 
Thus $\hat{Q}_m|v\rangle=0$. It means that $\langle v=0|\hat{Q}_m\,\hat{\rho}_m(0)|v=0 \rangle=0$.
On the other hand, when the monopole condenses, 
the vacuum is not eigenstate of $\hat{Q}_m$. Thus, $\hat{Q}_m|v\rangle\neq 0$. 
It means that $\langle v|\hat{Q}_m\,\hat{\rho}_m(0)|v \rangle \neq 0$.
Therefore, we find that the chiral non symmetric pair production takes place in the monopole condensed vacuum.

We have recently explicitly shown\cite{iwazaki3} that the chiral condensate $\langle\bar{q}q\rangle \neq 0$ 
arises when the monopole condensation $\langle \Phi\rangle \neq 0 $ takes place; $q$ denotes light quarks u or d.
The result has been shown by using effective monopole quark interactions describing the boundary conditions of
the chiral symmetry for the quarks.

%
%
%
%
%
%
%
%
%
%

\end{document}